# Perspective on first-principles studies of 2D materials

Asha Yadav[1], Carlos Mera Acosta[2], Gustavo M. Dalpian[2]*, Oleksandr I. Malyi[1]*

[1]ENSEMBLE[3] Centre of Excellence, Wolczynska 133, 01-919 Warsaw, Poland

[2]Centro de Ciências Naturais e Humanas, Universidade Federal do ABC, 09210-580 Santo André, São Paulo, Brazil

**Emails:** gustavo.dalpian@ufabc.edu.br (G. D), oleksandrmalyi@gmail.com (O.I.M)

## Abstract

The successful exfoliation of graphene from graphite has brought significant attention to predicting new two-dimensional (2D) materials that can be realized experimentally. As a consequence, first-principles studies of novel 2D materials become a routine, with thousands of papers published every year. What makes these studies interesting is that they predict new materials which have not been realized yet but should be a panacea for topological insulators, next-generation battery materials, novel solar cell materials, etc. There is no doubt that some of the proposed materials can provide a specific solution and their properties/performance can be confirmed experimentally, at the same time there are many false predictions because of the computational errors or the Lego-land approach to study 2D materials. To reduce the gap between theoretical and experimental works, we perform a systematic review of computational and Lego-land factors that should be minimized in future theoretical works.

## 1. Introduction

With the development of nanotechnology and the reduction of the dimensionality of the compounds, two-dimensional (2D) materials – nanosystems having one of the dimensions significantly smaller than others have been discovered. This field has become especially prominent with the successful exfoliation of single-layer graphene from its bulk counterpart (i.e., graphite) using a scotch-tape method.[1] Since that time, at least in the theoretical literature, it has been common to refer 2D compounds as single-freestanding 2D layers. However, while some compounds can indeed be exfoliated as a single freestanding layer (e.g., phosphorene[2,3], h-BN[4,5], and $MoS_2$[6]) many others can only be made 2D when grown on a substrate or having a very large thickness (e.g., silicene can only be stabilized on a specific substrate, and all attempts to stabilize it as freestanding materials have been unsuccessful[7]). Such experimental results imply that it is likely that the formation of freestanding vs on substrate 2D materials can be understood from the intrinsic properties of 2D materials (like energetics) as the formation of a 2D material require breaking weak interlayer van der Waals (vdW) (rarely chemical) bonds in the bulk compound to separate a single layer. For 2D materials on a substrate, the latter can act as a stabilizing agent, often results in chemical bonding between the substrate and 2D material[8]. Hence, if the isolation energy of 2D material is larger than the threshold value, it is unlikely that the compound can be realized in a freestanding form (discussed below).

What makes 2D materials exciting is that dimension reduction results in intrinsic anisotropy, quantum confinement effect, and exotic exciton physics that are not available in their bulk counterparts (discussed below). Additionally, owing to the vdW nature of the interaction between the layers, the properties of 2D materials can be controlled by tuning interlayer stacking[9] or combining different 2D materials in the heterostructures.[10] Since the experimental synthesis of 2D materials is largely based on a trial-and-error approach (i.e., the Edisonian approach), the first-principles studies have entered the field with investigations of already synthesized materials and theoretical predictions of hundreds of new 2D materials and their properties. Such works already suggested many potential new 2D compounds and even led to the development of general-purpose open-access databases such as 2DMatPedia[11], Computational 2D materials database[12,13] (C2DB), Materials Cloud two-dimensional crystals database[14], and MaterialsWeb[15] and also property-specific high throughput collections for spin-splittings[16] and magnetic properties[17]. Even with this kind of organization, the theoretical/computational literature in the field is somewhat chaotic. For instance, one often can question if the studied 2D materials are experimentally realizable or if sufficiently accurate methods have been used to predict the compounds. While the above questions can be related not only specifically to 2D materials,[18,19] the case of 2D materials is somewhat more special. First, for bulk three-dimensional (3D) materials, there are already well-defined experimental databases with crystal structures as verified with different methods for structural analysis, and the stability criteria are somewhat known (e.g., see, for instance, OQMD[20,21], Materials Project[22], and AFLOW[23]). However, for 2D materials, despite some ground-breaking work, the majority of current theoretical investigations are still focused on "created by hand" 2D compounds, which often do not have bulk analogous and for which realizability is never discussed. Moreover, the description of 2D materials often requires proper treatment of vdW forces which cannot be done based on the most widely used exchange-correlation (XC) functionals (e.g., Perdew–Burke-Ernzerhof (PBE)[24]). Finally, the properties of 2D materials in a plane-wave approach are described with periodic boundary conditions - meaning that the 2D materials are modeled with the periodic cell having an introduced vacuum in one direction (see calculations of optical properties described below). All these, thus, results in questioning how the properties of the materials should be investigated and if one can simply adapt the methods used for the bulk compounds to the 2D materials for the analysis. Motivated by this, herein, we provide a comprehensive overview of the recent progress in the first-principles investigation of 2D materials, demonstrating the above questions and suggesting best practices for accurate descriptions of the properties of 2D materials. In Sec. II, we will describe some key properties of 2D materials and how they should be described. In Sec. III., we will overview recent progress on predicting novel 2D materials and their properties. The summary and perspective for density functional theory (DFT) studies of 2D materials are given in Sec. IV.

## 2. Exciting materials properties enabled in 2D systems and best first-principles practices to describe them

**2.1. New features in electronic properties of 2D materials:** Formation of 2D materials results in bond breaking (usually weak vdW or rarely chemical bonds), which can lead to a totally new electronic structure, unlike their bulk counterparts. For instance, bulk $MoS_2$ (space group (SG): P6$_3$/mmc) is an insulator with indirect band gap energy of 0.91 eV with valence band maximum (VBM) located at Γ while conduction band minimum (CBM) between Γ and K as shown in Fig. 1a, while its monolayer is a direct band gap insulator with the band gap energy of 1.9 eV as calculated by using DFT-D2[25] (see Fig. 1b). Such a change of electronic properties is caused by kinetic-energy controlled quantum confinement and potential-energy controlled band localization/repulsion[26] and is a common feature of the 2D materials. For instance, the band gap energy of phosphorene changes from 1.46-1.59 to 0.33-0.87 eV as the number of layers increases from one to five layers (Fig. 1c), which is consistent with experimental trends.[27]

***Common problems in the description of fundamental properties of vdW systems***: While the above results demonstrate the examples of successful description of experimentally observed electronic properties of 2D materials within the first-principles calculations, most widely used XC functionals have some fundamental limitations in their description: (i) similar to bulk materials, an accurate description of the electronic structure requires use of XC functional which has reduced self-interaction error; [28] (ii) modern XC functionals (e.g., PBEsol[29] and strongly constrained and appropriately normed (SCAN)[30] meta-GGA) and even hybrid functionals (e.g., Heyd–Scuseria–Ernzerhof (HSE) functional[31]) are still unable to describe the vdW forces accurately; (iii) Recently, developed methods for accounting of vdW forces are based on the soft XC functionals (deviating from linear Koopman's condition[32]), which thus suggest that there is no single unified method that is able to describe both structural and electronic properties of 2D materials with high accuracy. To better illustrate points (i-iii), let us consider the example of bilayer phosphorene, which has weak interlayer bonding. When the PBE XC functional is used for both structural relaxation and the band gap calculations, one finds that the band gap energy is 0.29 eV with the interlayer distance is 3.50 Å which is significantly different from the HSE band gap energy of 1.22 eV calculated for the same structure (i.e., one obtained from PBE relaxation)[33]. Importantly, using PBE functional results in significant overestimation of interlayer distance as compared to the methods able to accurately describe vdW forces, which is essential for calculations of band gap energy (Fig. 1d). For instance, interlayer distance in bilayer phosphorene is 3.50 and 3.09

Å with PBE and optB88-vdW XC functionals, respectively[33]. One may naively think that the treatment problem of vdW systems refers only to multilayer systems. However, the band gap energy of monolayer phosphorene is sensitive to the treatment of vdW forces. Thus, the HSE band gap energy for the PBE relaxed phosphorene structure is 1.62 eV, while the corresponding value for structure relaxed with optB86b-vdW[34] functional is 1.46 eV. This is not surprising as the phosphorene has a buckled honeycomb lattice structure with coexisting low and high electron density regions for which, accounting non-local correlation is vital for structural relaxation.

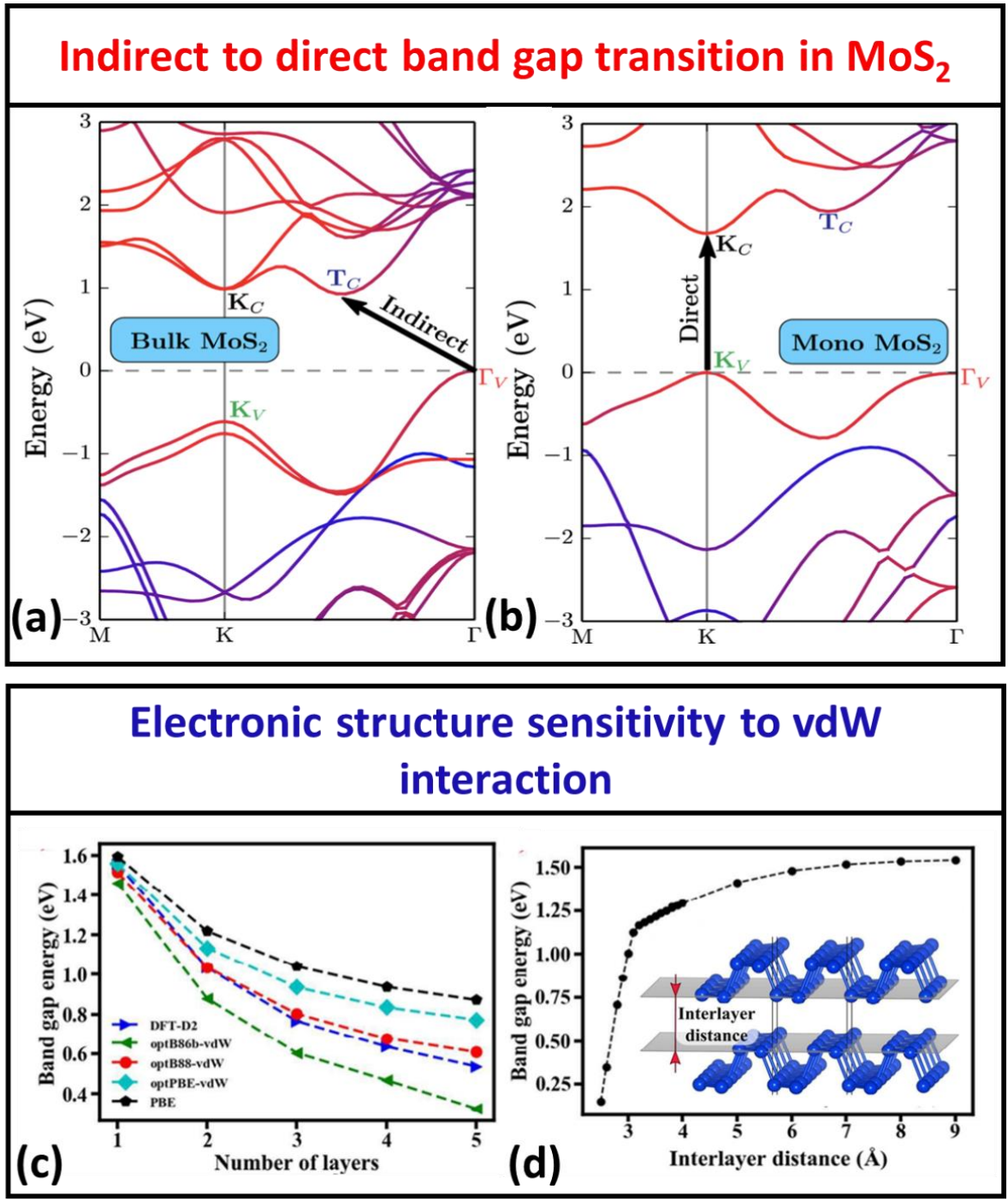

**Figure 1.** Atom projected electronic band structure of $MoS_2$ in (a) bulk and (b) monolayer form. The blue and red colored bands in the band structure represent S and Mo projections, respectively. Fig. (a) and (b) are adapted with permission from ref[26] Copyright (2015) American Chemical Society. HSE electronic band gap calculated as a function of (c) the number of layers using different XC functionals for internal relaxation of bilayer phosphorene and (d) as a function of interlayer distance. The inset in Fig.(d) represents the interlayer distance in bilayer phosphorene. Fig (c) and (d) are reproduced from Ref.[33] with permission from the Royal Society of Chemistry.

***What is the best practice for first-principles calculation of vdW systems***: Taking into account the above limitations of the existing DFT methodology, the internal structure of the compound should be described with the XC functionals that, in general, are able to capture the vdW forces accurately. While the electronic structure should be analyzed for structures relaxed using vdW corrected functional with accurate methods to describe electronic structure (e.g., hybrid functional[31] and GW[35]). However, the open question is what is the most accurate way to describe vdW bonded systems as there is a range of vdW functionals already developed. For instance, the DFT-D approach[25,36,37] has been developed as a way to add an empirical correction (i.e., $E_{disp} = -s_6 \sum_{i=1}^{N_{at}-1} \sum_{j=i+1}^{N_{at}} \frac{C_{6ij}}{R_{ij}^6} f_{dmp}(R_{ij})$, where $s_6$ is the global scaling factor and summations indicate the sum over all $N_{at}$ atoms of the unit cell, $E_{disp}$ and $C_{6ij}$ are the vdW-dispersion corrected energy term and dispersion coefficient corresponding to pair of atoms i and j, respectively, $f_{dmp}(R_{ij})$ is the damping function used to account near-singularity value of $E_{disp}$ for a small value of $R_{ij}$ (interatomic distance between pair of atoms I and j) that scales the interaction between atoms i and j) on top of already widely used functional. This method has been widely used to model 2D materials and often can predict relatively good interlayer distance. However, it has a significant drawback as the $C_{6ij}$ coefficient is independent of the nature of chemical bonding (at least in the first versions of DFT-D), and consequently, the empirical correction by itself does not directly account for the charge density distribution in the compound. This thus suggests that often one can get the correct interlayer distance but for the wrong reason. The second type of approach has been developed via including non-local correlation term to the XC functional[38-44] (i.e., $E_c^{nl} = \frac{1}{2} \int d^3r d^3r' n(r) \Phi_0(r,r') n(r')$, where $E_c^{nl}$, $n(r)$ and $\Phi_0(r,r')$ are non-local correlation energy correction, total electron density and correlation kernel). The main fundamental advantage of this approach over DFT-D is that the non-local correlation term is electron density-dependent[45] and, in principle, allows to account for charge-density rearrangements caused by vdW interaction. So far, a range of vdW-DF functionals have been developed including vdW-DF0[38-41], vdW-DF1[42,43], vdW-DF2[44], (opt-PBE-vdW, optB88-vdW, optB86b-vdW)[34,46] and vdW-DF-cx[47]. All these functionals have some advantages and disadvantages. However, more recent comparison of vdW-DF results with diffusion quantum Monte-Carlo indicates that there is no single treatment of vdW forces at the first-principles level that allows to correctly reproduce charge-density distribution in bilayer phosphorene[48]. Taking into account these results, from our perspective, one should still use vdW-DF functional (unless more advanced methods are developed) but make sure that the predicted results are not highly sensitive to the choice of the vdW-DF functional. If the results become too sensitive to the different treatments of vdW forces, then one should honestly admit that first-principles calculations do not have the prediction power needed to describe the corresponding effect.

## 2.2. Optical/dielectric properties of 2D materials:

Similar to the case of bulk materials, the optical properties of 2D materials are one of the most fundamental properties. What makes 2D materials unique is the strong excitonic effects-Coulomb interaction between electron excited to the conduction band and hole present in the valence band (Fig. 2a) modifies the optical properties of 2D systems drastically. Theoretical studies predict the exciton binding energy of about 0.5-1 eV for 2D transition metal dichalcogenides[49], which is comparable with recent experimental results.[50] In particular, for phosphorene and 2D-MoSe₂, the exciton binding energies determined as differences of electronic and optical band gaps (see Fig. 2b) were reported to be around 0.9[51] and 0.55 eV[52], respectively. This energy is big enough for exciton observation even at room temperature, allowing to design fundamentally new devices utilizing excitonic effects.[50] Taking into account electron-hole interaction is highly sensitive to the screening environment, it is possible to control the excitons by using different substrates, encapsulation, etc., opening the possibility for their programming.[53,54]

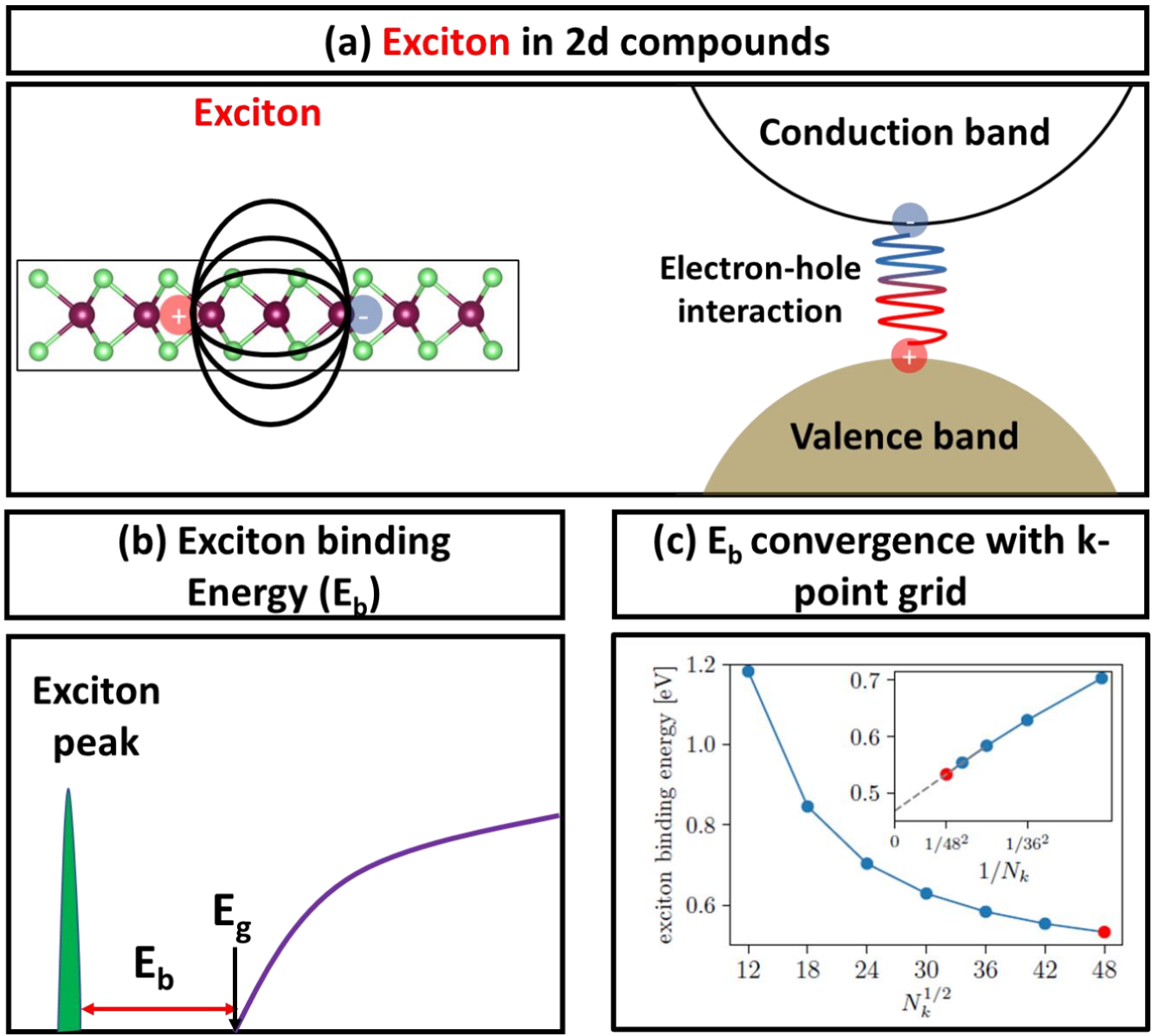

**Figure 2.** Schematic representation of (a) exciton in 2D material and (b) exciton binding energy. (c) Convergence of exciton binding energy calculated using the Bethe-Salpeter equation in a monolayer of MoS₂ with respect to k-point grid. The red point in the plot and inset in (c) shows k-point sampling overestimating exciton binding energy from the extrapolated value by 0.06 eV. Fig (c) is adapted from ref. [12] and used under the basis of Creative Commons License.

***Common problems in the description of optoelectronic/dielectric properties of 2D materials***: For bulk compounds, dielectric properties of a given compound can be described with sufficient accuracy using, for instance, density functional perturbation theory[55–58] or/and independent particle approximation.[59,60] However, both methods cannot account for the excitonic effect, and hence the Bethe-Salpeter equation[61–64] should be used. Second, both methods have been developed for bulk materials where the materials properties (e.g., volume) are defined by atomic identities, composition, and structure (Fig. 3a). However, within periodic boundary conditions, the 2D material is described as a slab with a vacuum of large thickness (Fig. 3b) where the amount of vacuum directly affects results. An example of imaginary part of dielectric function is shown in Fig. 3c for the case of a single $MoS_2$ layer, where the variation of vacuum thickness results in the vacuum volume-dependent dielectric function (and hence absorption spectra, not shown here). Such naive application of the DFT theory towards the calculation of optical properties is indeed a common mistake in the field, and rarely optical calculations are carried out correctly.

***What is the best practice for first-principles calculations***: We strongly believe that one should never use/present dielectric function data for 2D materials predicted without the volume correction as such data can mislead others and lead to unphysical conclusions (e.g., unphysical device performance) at least for the cases when the focus is on optical properties. The question remains however how to properly account for such correction. Generally, the macroscopic dielectric function of a material generated as a result of external plane-wave $V_0 e^{iq.r}$ potential can be given as[65]

$$\frac{1}{\varepsilon_{M(q)}} = \frac{<\widetilde{V_q}>_\Omega}{V_0} = \varepsilon_{GG'}^{-1}(q) = \varepsilon_{00}^{-1}(q)$$

where G, q, $\varepsilon_{M(q)}$, and $\varepsilon_{00}^{-1}(q)$ are reciprocal lattice vector, wave vector in the first Brillouin zone, macroscopic, and microscopic dielectric function, respectively. $<\widetilde{V_q}>_\Omega$ is the lattice periodic function averaged over a unit cell. However, generalization of $\varepsilon_{M(q)}$ to low dimensional systems like 2D materials leads to potential averaging over the whole vacuum plus slab region. Thus, extending bulk methods to 2D material will be misleading. A different way of averaging over the slab region is proposed in ref [65] as

$$\frac{1}{\varepsilon_M^{2d}(q_{||})} = \frac{2}{d}\sum_{G_\perp} e^{iG_\perp z_0} \frac{\sin\left(G_\perp \frac{d}{2}\right)}{G_\perp} \varepsilon_{G0}^{-1}(q_{||})$$

where $z_0$ is the centre of the slab with the thickness $d$ of the slab. $q_{||}$ corresponds to $q$ confined to the plane of the slab. The summation is carried over all the $G$ components with $G_{||} = 0$. This method calculates macroscopic dielectric function by accounting contributions from only slab part excluding vacuum. In addition to this, another alternative method to incorporate volume correction is to use the volume of the slab excluding the vacuum part ($\Omega_{2D}$, e.g., average volume per layer in the bulk form of the material) instead of the volume of the whole cell containing 2D materials (i.e., $\Omega_{3D})$[66] as shown in the Fig. 3b,3d. In this way, one first needs to define the thickness of 2D materials and then calculate the corresponding imaginary part of the dielectric function. This simple approximation allows to solve the problem of dielectric function dependence on vacuum thickness. We note, however, that this method should be used with caution especially when different definitions of the thickness of 2D materials affect the main conclusions. Another important part of the calculation of the optical properties involves accounting for the role of excitons, which can be done using the Bethe-Salpeter equation.[61–64] From a practical point of view, one should remember that exciton binding energy is highly sensitive to k-point density (Fig. 2c), hence in principle, improper selection of k-points may have a dramatic influence on optical property due to its non-analytic nature at a lower density of k points.

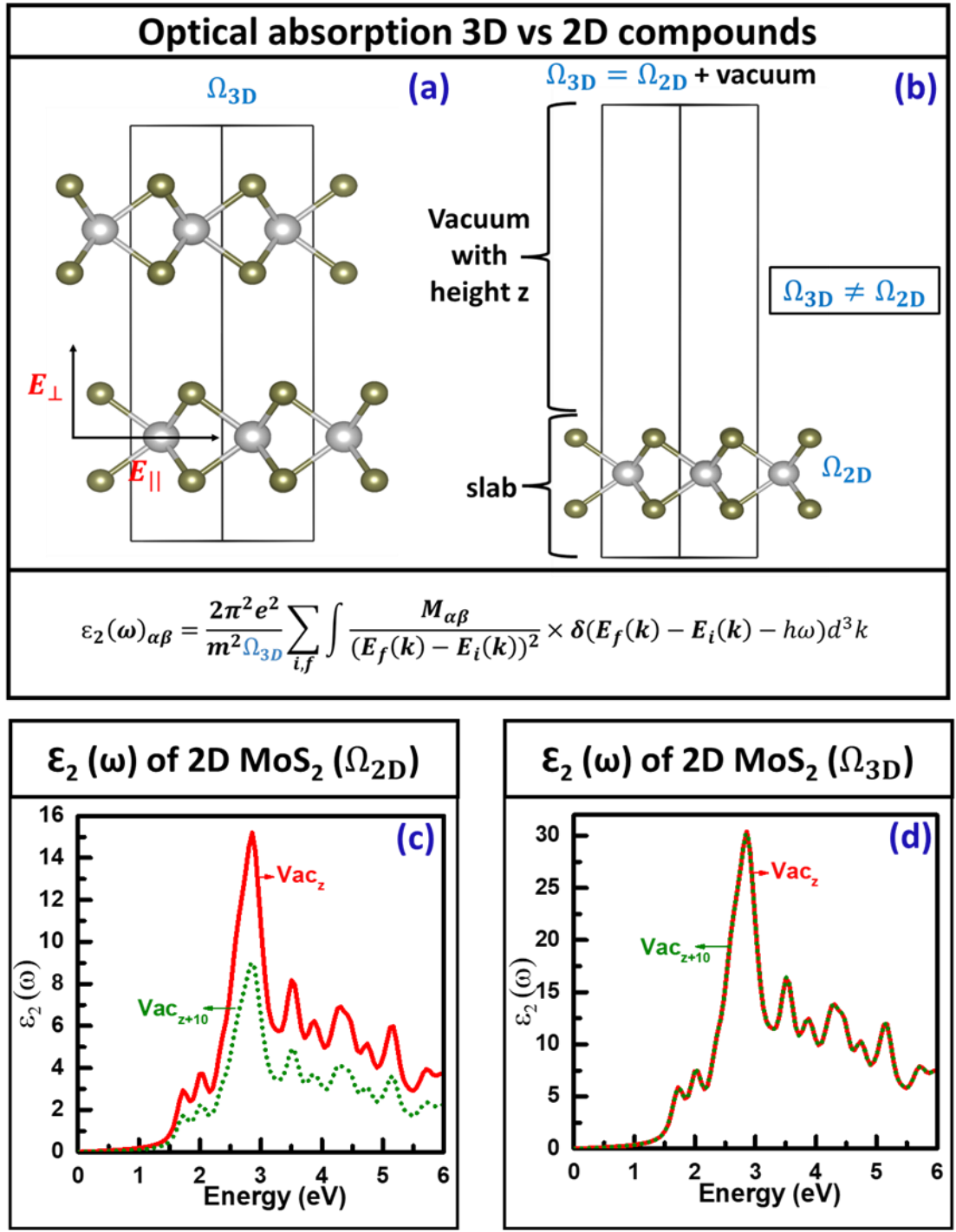

**Figure 3.** Side views of (a) bulk $MoS_2$ (b) monolayer $MoS_2$. Here, $E_\parallel$ and $E_\perp$ are the parallel and perpendicular components of electric field. Imaginary part of the dielectric function ($\varepsilon_2(\omega)$) of $MoS_2$ calculated using independent particle approximation for different vacuum thicknesses (c) without including volume correction (d) with including volume correction. In the equation of $\varepsilon_2(\omega)$, $\Omega$, $\omega$, h, e, and m are cell volume, angular frequency, Plank's constant, charge and mass of the electron, respectively. $E_i(k)$ is the energy level for a given wavevector k. Here, i and f correspond to initial and final state, respectively. $M_{\alpha\beta}$ is a square matrix with momentum elements in crystal directions α and β. Volume correction

in Fig (d) is accounted by multiplying $\varepsilon_2(\omega)$, with $\Omega_{3D}$ and then normalized by $\Omega_{2D}$ (calculated by assuming that each monolayer layer in the cell occupies a space equal to the van der Waals distance between layers).

## 3. Prediction of novel 2D materials and their properties

### 3.1. General strategies for the search of novel 2D materials

Since the discovery of graphene, an enormous number of works have been published on predicting novel 2D materials. While the initial works were mainly focused on the investigation of graphene analogs (e.g., silicene, phosphorene, and h-BN), more recently, the rise of the area of materials discovery and design has opened a broad avenue for the prediction and development of new, undiscovered 2D materials. These include the development of dedicated 2D materials databases[11–17] where thousands of different compounds can be found, together with many of their calculated properties, and the development of theoretical/computational strategies for predicting new, not yet synthesized materials. One can classify these strategies into two different fashions, similar to what is done in nanostructures: bottom-up or top-down approach. In the bottom-up approach, the idea is to substitute the chemical elements of a known compound (e.g., $MoS_2$) by other elements with similar electronic structures, creating compounds such as $MoTe_2$ or $WS_2$. In the top-down approach, one starts from bulk materials and looks for compounds that could be, in principle, exfoliable. Different algorithms used to explore 2D materials from its bulk counter part are discussed below.

The *first-generation algorithms* for searching of exfoliable 2D materials involve analysis of the stacking of atomic planes in some specific directions in the unit cell of a bulk compound. This can be either the *c* lattice vector, which is usually the longer one or a set of directions in which the algorithm loops. The simpler way is to use a linear combination of lattice vectors. These algorithms search for a plane which can be separated as a monolayer from its bulk counterpart, i.e., exfoliate, accounting for chemical bonding among these planes through determined criteria, e.g., the sum of covalent or vdW atomic radius plus a tolerance factor. Using this kind of methodology, one can easily find exfoliable materials that belong to similar classes as graphene. However, this method can easily miss materials that are buckled or with some rough patterns inside their crystal structure. To solve this problem, a second-generation *algorithm* was created - Instead of evaluating atomic planes along fixed directions, these methods analyze the network of connections that form a bonded cluster in the bulk material. Once these clusters are identified by measuring the distance among atoms, the existence of layered patterns is determined by one of the two methodologies: topology-scaling algorithms (TSA)[15] or rank determination algorithms (RDA)[14]. TSA takes into account the rate of how the bonded cluster grows when the cell uniformly increases in size. For instance, an original bonded cluster in a structure contains $N_1$ atoms. The same cluster inside a supercell of n×n×n unit cells will contain $N_2$ atoms. By analyzing the ratio of $N_2/N_1$, it is possible to determine the dimensionality (d) of such cluster by expecting this ratio. The RDA algorithm also analyzes bonded clusters embedded inside supercells, but the dimensionality is determined by the rank of the matrix composed by the coordinates of a given atom and its periodic images that belong to the same cluster. For instance, if a given cluster of atoms is set to form a 2D monolayer, a set of just two linear independent vectors can be chosen to map any atom coordinate to its periodic images in the same cluster, disregarding the orientation of such cluster in Cartesian space. Such methodologies can automatically account for rough patterns inside the crystal. However, this kind of methodology still has some drawbacks since it relies on fixed criteria for finding atoms that are bonded together, forming the clusters.

The third generation of algorithms relaxes the constraint of using a fixed parameter to determine the dimensionality of a specific compound.[67] The concept involves the determination of a scoring parameter that will indicate the degree of dimensionality (the degree of "1D-ness," "2D-ness," etc.) of the material. These different strategies have been applied to 3D materials databases such as the Inorganic Crystal Structure Database (ICSD)[68] and the Crystallography Open Database[69] (COD), ranking the materials according to their degree of dimensionality. The different algorithms described above provide a powerful basis for the discovery and design of new 2D materials, leading to the construction of large materials databases as described in the section below. As these methods only account for the crystal structure of materials, straightforward questions can be raised regarding their stability and synthesizability.

## 3.2. Different 2D materials databases

The advent of computational repositories significantly changed the way the computational materials science area advances. The basic properties of all synthesized inorganic materials have already been calculated and reported,[20,22,23] including in many cases band structures, formation energies, phase diagrams, and even some stress tensors. The popularity of these inorganic materials databases has opened opportunities for other, smaller, databases focused on materials with specific properties, including databases on 2D materials such as the 2DMatPedia[11], C2DB[12,13], Materials Cloud two-dimensional crystals database[14] and MaterialsWeb[15] to mention a few. These databases mainly followed some type of high-throughput or machine learning (ML) approach (discussed below) that reported the properties for thousands of unique 2D entries, some of which have been synthesized and the majority of which are new theoretical predictions. For instance, the C2DB reported around 4000 unique entries, all generated compounds follow the strategy shown in Figure 4a. These hypothetical materials then pass through a complex automated workflow (Figure 4b) where ab initio calculations are performed, and materials properties are calculated. In this step, the stability of materials is evaluated through the formation enthalpies and the simplified convex hull (i.e., using bulk OQMD data directly compared to corresponding energies of 2D materials in the C2DB). The development of these 2D materials databases certainly represents a major advance towards the design and discovery of new materials. However, even though the advances have been enormous during the last few years, a few challenges still remain. These challenges will be discussed below.

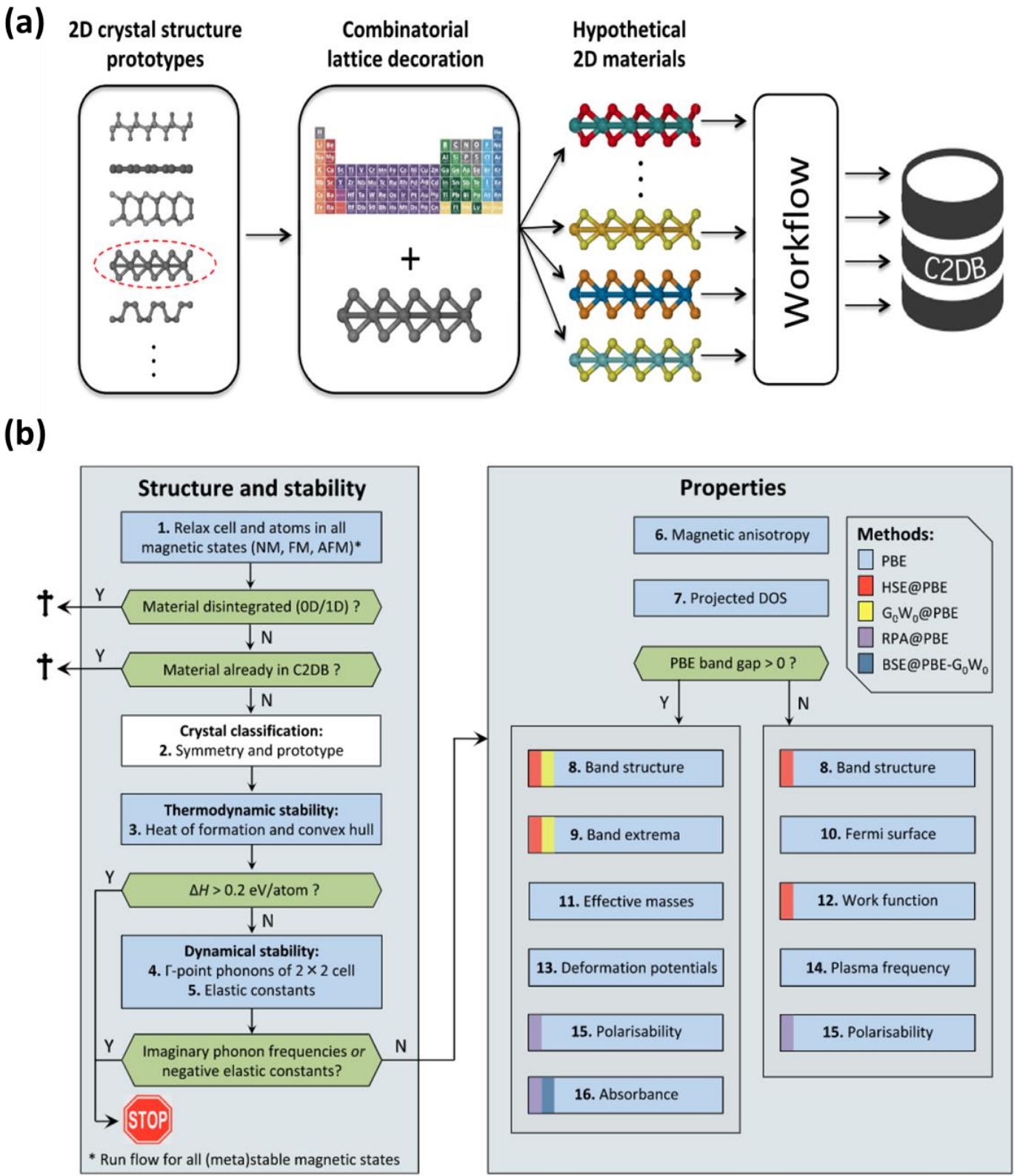

**Figure 4:** (a) General strategy used to generate entries in the C2DB database. (b) Detailed description of the C2DB workflow. Fig (a) and (b) extracted from ref[12] and used under the basis of Creative Commons License.

### 3.3. ML prediction for 2D materials

ML has been intensively explored in recent years and extended to all possible areas of materials sciences and condensed matter, naturally, including study of 2D materials. In this context, the development, application, and utility of ML techniques are usually focused on (i) the prediction of novel 2D material prototypes (Fig. 5),[70] (ii) the design of compounds with target properties such as ferroelectricity, topology, and magnetic order[71–73], and (iii) the regression and classification of materials properties such as bandgap, conductivity, topology, and spin-polarization[74]. The current stage of the ML application to 2D materials also includes the study of external perturbations such as strain and electric field and patterns of properties under the compositional and structural changes[71,75]. In this section, we cover the recent ML progress in 2D materials, which can be divided into two big categories (i) materials prediction and (ii) prediction and optimization of materials properties.

***Prediction of novel 2D materials:*** The early works of ML for the prediction of 2D materials are based on the global optimization approach.[76] Examples include the genetic algorithm for structure searching combined with DFT calculations to identify low-energy 2D structures of group-IV dioxides $AO_2$ ($A$=Si, Ge, Sn, Pb).[77] Bahmann and Kortus developed an evolutionary algorithm that can search for 2D crystals[78], and Zhou et al.[79] extended an evolutionary algorithm to search for 2D structures. Both algorithms help to search for structures with fixed stoichiometry and the number of atoms per cell. Luo et al. extended a particle swarm optimization algorithm to search for fixed stoichiometry 2D structures that are both completely planar[80] and have finite-layer thicknesses.[81] Recently, Song et al.[82], proposed a deep learning generative model for composition generation combined with random forest-based 2D materials classifier to discover new hypothetical 2D materials, also using element substitution structure to predict newly predicted hypothetical formulas. The authors reported 267,489 new potential 2D materials compositions and confirmed twelve 2D layered materials by DFT formation energy calculation. With the goal of discovering novel 2D materials with unusual compositions and structures not limited or based only on pre-existing prototypes. Revard et al. developed a grand-canonical evolutionary algorithm that searches the structural and compositional space while constraining the thickness of the structures.[83]

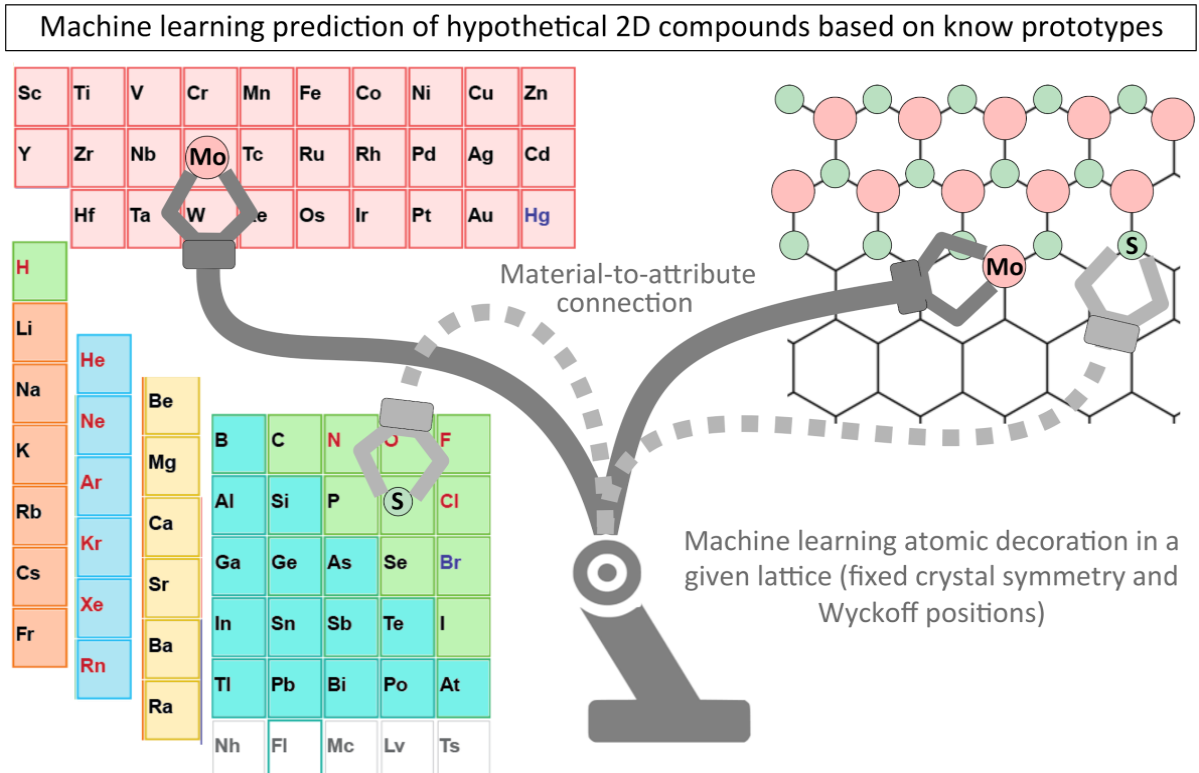

**Figure 5.** Schematic diagram to illustrate machine learning method to generate a number of 2D compounds having fixed crystal symmetry and Wyckoff positions using a bottom-up approach. This fig is inspired by ref.[73]

Besides the theoretical prediction of hypothetical 2D compounds based on the genetic algorithm and neural networks, another strategy is focused on the direct prediction of the target properties that define the potential existence of a 2D materials candidate, e.g., the energy above the convex hull, formation energy, decomposition temperature, phonon spectrum, and melting temperature. This allows for directly predicting the thermodynamic stability with respect to competitive phases and dynamic stability. For instance, Mortazavi et al. proposed a machine-learning model for interatomic potentials trained with computationally *ab-initio* molecular dynamics trajectories that in contrast to DFT simulations eliminates the nonphysical imaginary frequencies in the phonon dispersion curves.[84] Using this method, the authors calculated the phonon dispersion and dynamical stability of novel 2D compounds. To generate a predictive ML model for thermodynamic stability, Schleder et al. classified compounds into high, medium and low stabilities based on DFT-calculated formation energy and energy above the convex hull.[70]

***Prediction of 2D materials with target properties***: As previously described, sometimes one wishes to predict compounds with a specific target property aiming for potential applications. The ML then not only gives a reasonable fitting for the target functionality but also allows for predicting novel 2D crystal symmetries and atomic combinations that optimize the target property. Thus, the ML algorithm aims to learn specific functionalities that are not necessarily related to compound stability; however, it is naturally desirable to predict compounds with a given functionality that are also thermodynamically and dynamically stable. This is not always the case, since the complete verification of the stability in 2D compounds is a complex (but necessary) task[85], as previously discussed. The ML prediction of compounds with target functionalities includes the discovery of potential 2D ferroelectric metals[71] , 2D hybrid perovskite materials with large bandgaps[86], the topological order in 2D toy models[87] as well as the prediction of 2D quantum spin Hall insulators[88], 2D magnets[89], optoelectronic transitions.[90] The works that involve the prediction of materials using training data calculated using approximations (e.g., DFT, quantum monte Carlo, or molecular dynamics), will always implicitly be the bias product of these approximations.This fact must always be considered since the approximations  are made to learn theoretical relationships and that the exotic effects predicted theoretically within a set of approximations need not exist in normal environmental conditions.

### 3.4. Limitations of the majority of current 2D works

Despite the enormous progress that has been carried out on the first-principles exploration of novel 2D materials, there is still a question regarding the realization of theoretical predictions. In fact, the majority of 2D works still report properties of 2D materials created by hands (Lego-land approach, e.g., the substitution of C atoms in graphene by Si[91] or using some experimentally reported structures to guess other 2D  materials[92] ). This eventually leads to the situation when many different groups report 2D materials having the same composition and totally different electronic properties (e.g., SiP[93–96]). Even the 2D materials databases reviewed above, despite their seminal nature, cannot provide sufficient information on materials realization. For 3D materials, we know (or at least assume) that prediction of the convex hull (e.g., full exploration of competing phases of compounds) and investigation of dynamic stability can provide some information on materials realization. Thus, if the compound is located above the convex hull - tends to decompose, it is unlikely to be realized if its decomposition energy is lower than some threshold value.[97–99] For 2D materials, however, despite some works on the realization criteria (including the open-access databases), there are still no universally accepted guidelines to describe the realizability of predicted 2D materials. So let us discuss some of the typical criteria used in the literature.

**Phonon stability:** The absence of imaginary (negative) phonons is the most widely used criterion to analyze the "stability" of 2D materials in the DFT literature. Here, the presence of imaginary frequencies indicates that the energy of the system can be lowered by a set of specific atomic displacements, and hence it cannot be realized experimentally. To demonstrate the power of this method, we consider the example of 2D-NbS$_2$, which is predicted to be a phonon-unstable system, as shown in Fig. 6c,d. For the compounds which are phonon stable, there are no imaginary phonons, as shown on the example of 2D-MoSe$_2$ in Fig. 6a,b. Interestingly, that description of phonon spectra can be used not only to claim the phonon unstable systems but also to predict potential compounds having phonon stable spectra, as for instance has been demonstrated by the method of Stokes and Hatch.[100] For 2D materials , the phonon stability filter is the cheapest among the ones discussed in this work and the most widely used in the field. For instance, Mounet et al. provided phonon band structures for a large set of compounds in their open-access database.[14] However, in contrast to many other papers, we believe that the phonon filter does not allow us to discuss if the compound is dynamically stable or not, as strictly speaking phonon calculations only account for small atomic displacements without large structural rearrangements. Hence, while the phonon filter is important for predicting realizable 2D materials, it is only one of the criteria.

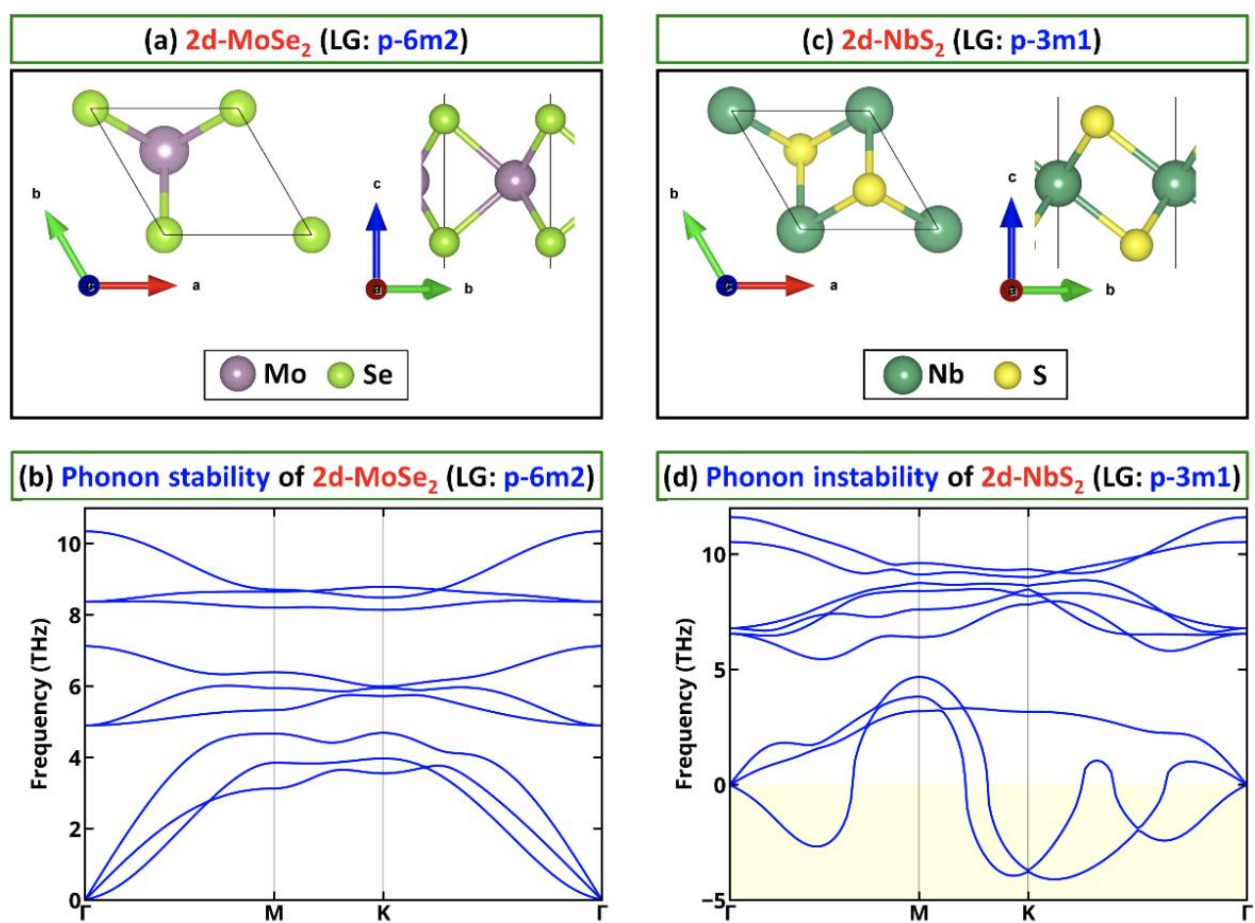

**Figure 6.** Application of phonon stability filter for 2D materials. (a) top and side view of 2D-MoSe$_2$ and (b) phonon band structure of stable materials having no imaginary frequencies. (c) top and side view of 2d-NbS$_2$ and (d) phonon band structure of unstable material having imaginary phonon (shown as negative phonon frequencies). Reprinted with permission from Ref.[85]. Copyright (2019) American Chemical Society.

**Dynamic stability:** In nature, the equilibrium state of the system is found by minimizing the Gibbs free energy at the corresponding temperature and pressure.[101] In computational solid state physics, such equilibrium state can be predicted by ab initio molecular dynamics (AIMD) simulations accounting for their specific limitations:

(i) *AIMD is limited in time of simulations:* Since the AIMD simulation is limited to a few hundred of picoseconds (ps), one cannot expect that the system will reach its equilibrium starting at a specific temperature within the typical simulation time. To overcome this problem, AIMD studies should be performed at a few different temperatures, and one should monitor energy and distribution of structural motifs to extract the evolution of the system. If the energy of the system can be reduced within a few ps at any temperature, it means that the system is unlikely to be realized experimentally. Fig. 7a demonstrates the example of the system where, at a low temperature (Fig. 7b), the system's energy is not lowered during the simulation time. However, AIMD simulation for the system at higher temperatures (Fig. 7c) results in substantial energy lowering suggesting that the compound cannot be realized as a 2D material.

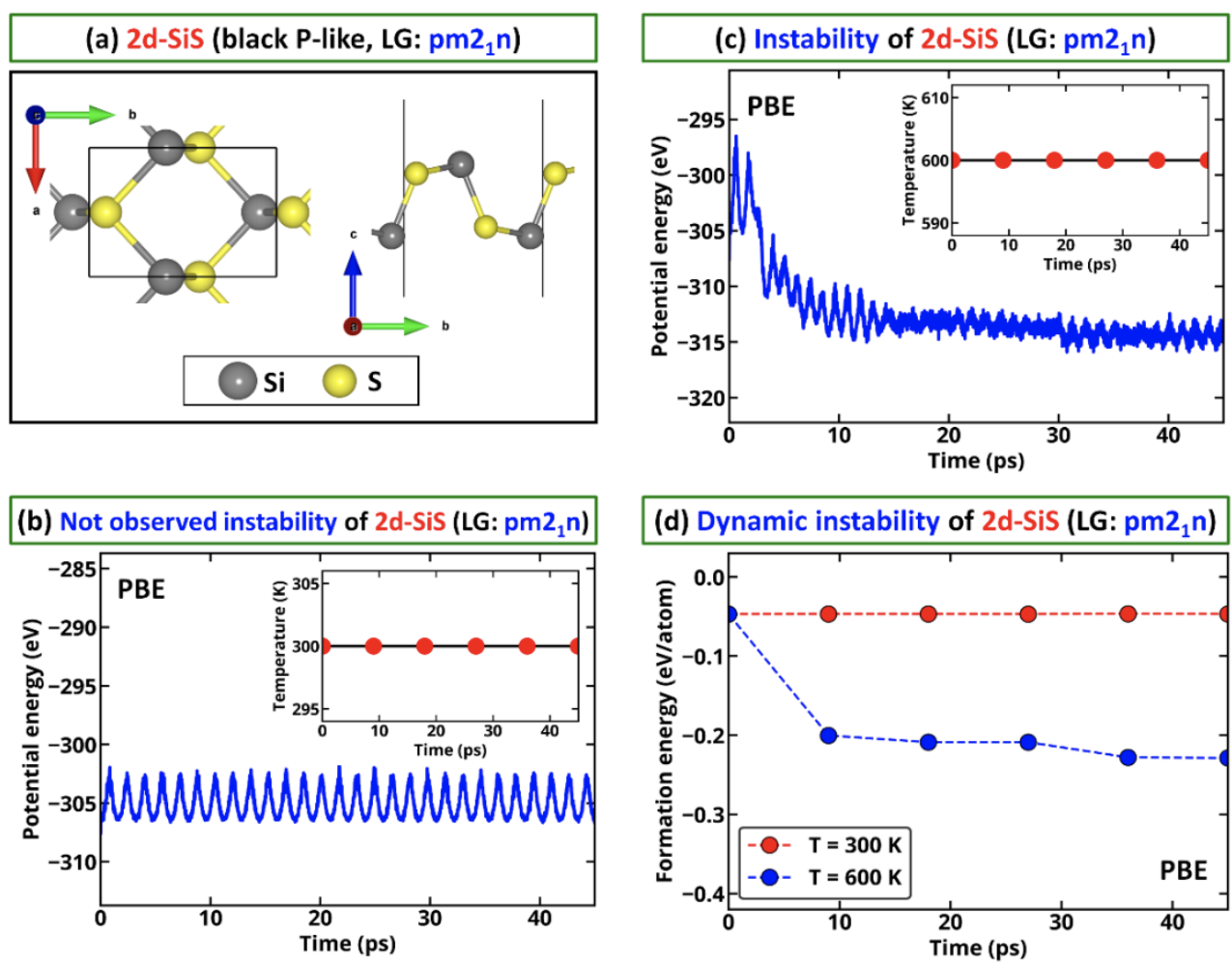

**Figure 7.** Temperature-dependent dynamical stability of 2D SiS system. (a) 2D SiS structure made from phosphorene structure (layer group pm2_1n) obtained by replacing P by Si and S atoms. Potential energy profile for AIMD simulation at (b) 300K (c) 600K. The insets of Fig. (b) and (c) provide the temperature protocol throughout the AIMD simulations. (d) Formation energy of the structure taken at various time steps of AIMD simulation. Reprinted with permission from Ref.[85]. Copyright (2019) American Chemical Society.

*(ii) Energy profile:* The potential energy of a system during an AIMD run with NVT (amount of substance (N), volume (V) and temperature (T) conserved) ensemble always fluctuates. This eventually can lead to the situation when energy lowering during AIMD runs cannot be visually seen from the potential energy profile. Indeed, the majority of modern AIMD simulations are initialized by some kind of random velocity distribution corresponding to the specific temperature - AIMD simulation profile for 2D materials usually includes both equilibration and production run at the same time. This eventually can lead to the situation when visually insignificant energy lowering symmetry breaking can be mistakenly ignored. Hence, to ensure the dynamic stability of 2D materials, one should explore the energetics of AIMD configurations (Fig. 7d). For instance, one can extract several snapshots and perform static DFT energy minimization for all of them. The proposed 2D material can be only dynamically stable if the energy of the configuration obtained at different time steps are not lower than that of the initial configuration.

**Energetic stability:** If a given compound satisfies the phonon and dynamic stability filters, one may wonder how much energy it costs to form the 2D material and if it can be done with the typical methods. For instance, Zhu et al.[102] stated that the proposed 2D material is stable if it has negative cohesive energy. Indeed, negative cohesive energy suggests that a compound does not decompose to individual atoms spontaneously, but it has rather limited meaning as most compounds have negative cohesive energy. Other works suggest that if the formation heat of a compound is negative, it is thermodynamically stable.[103] Obviously, if the formation heat of a compound is positive, it is unlikely that the compound can be synthesized. However, the negative formation heat is also not the true descriptor for the compound to be synthesized (see discussion in ref.[85]). For bulk systems, the realizability of the compounds is usually discussed in the language of the energy convex hull - where the compounds on the convex hull or close to the convex hull can potentially be synthesized. Indeed, the majority of the experimentally synthesized compounds have the energy above convex hull less than 100-150 meV/atom.[99]

Assuming that some instability (i.e., energy above convex hull) can be realized, one can estimate the threshold energy above which potential 2D materials can be considered as unstable by calculation of maximum energy above convex hull for already synthesized single layers of freestanding 2D compounds (Fig. 8). This threshold energy is expected to be different for different XC functionals (e.g., one accounting for vdW and not accounting for vdW correction). However, surprisingly within a small dataset of 2D materials, it turns out that the maximum value of energy above the convex hull is about 0.11-0.12 eV/atom, with a substantial difference only for PBE functional (i.e., 0.04 eV/atom). Application for such filter thus allows to disqualify some highly unstable materials (see discussion in ref.[85]). We note however that similar to bulk convex hull calculations, such analysis requires explicit accounting for the spin degree of freedom to predict a true convex hull. This is especially critical as with discovery of monolayer $CrI_3$ 2D magnet there is a growing interest in the 2D magnetic materials[73]. While state-of-the-art experimental measurements can reveal magnetic orders in 2D systems [104] , the magnetic configuration in potential 2D material is subject to investigation and not subject to guess or wishful choice. This thus requires detailed screening of different spin configurations and identifying the lowest energy configuration to predict the corresponding properties. Similar to bulk calculation, one should include identifying the lowest energy spin arrangement for different sizes of 2D supercell. We note that the lowest energy magnetic order can change from monolayer to multilayer system[105]. Such screening is critical not only to the analysis of materials stability but also to the prediction of basic materials properties. For instance, let us consider the case of 2D-CrI₃: if one studies the system properties using a non-magnetic configuration, DFT predicts that this material is metallic. Describing the same system for different magnetic configurations such as ferromagnetic or antiferromagnetic suggest material as a semiconductor with band gap 1.135 and 1.247 eV calculated using LDA XC functionals, respectively.[105]

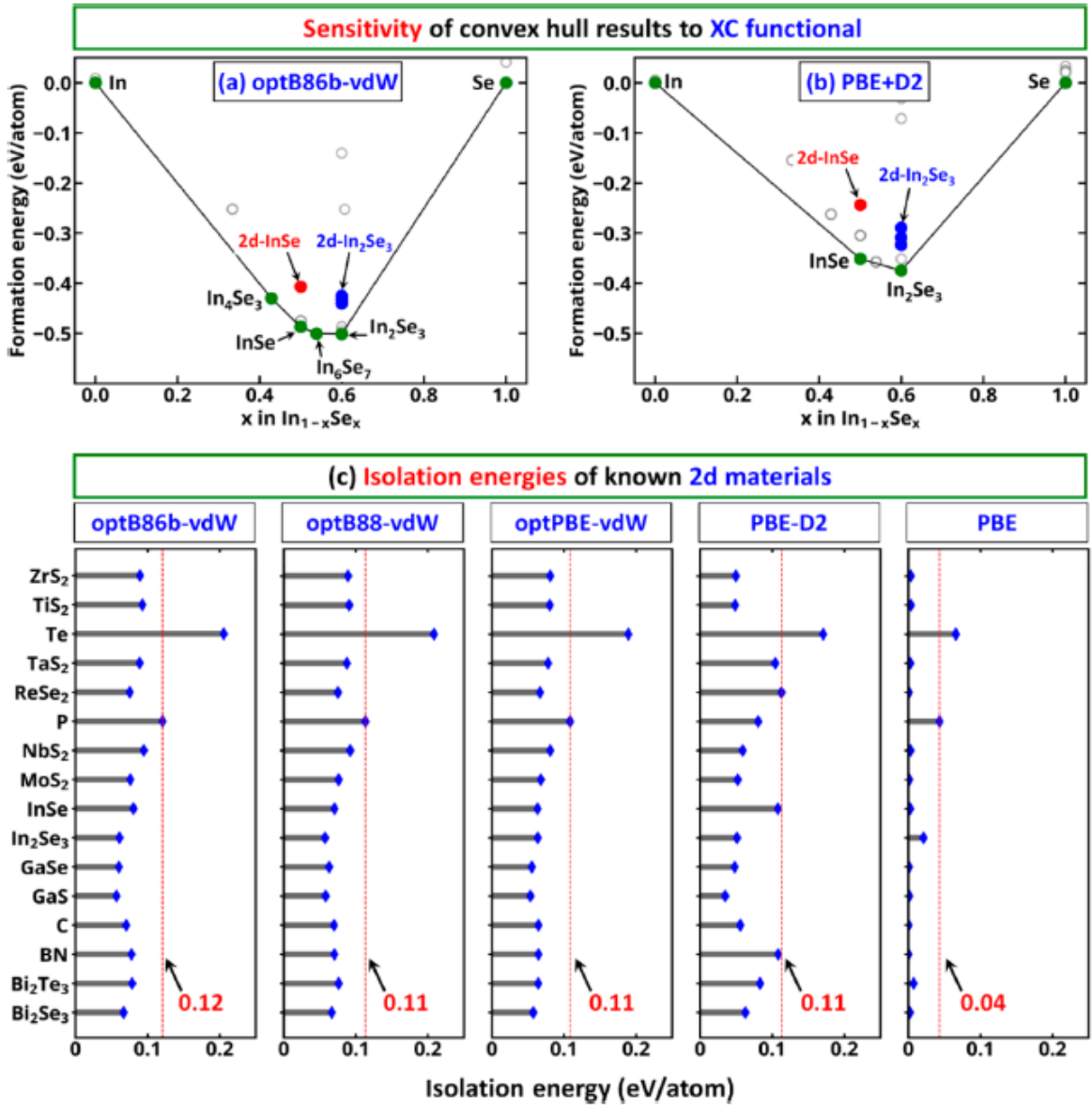

**Figure 8.** Sensitivity of convex hull energy of In-Se system towards (a) optB86b-vdW and (b) PBE-D2 XC functionals. The red and blue dots in Fig. (a) and (b) indicates the formation energies of 2D systems. (c) Isolation energy of different known compounds using various XC functionals. The dashed red line represents the expected threshold energy above the convex hull. Reprinted with permission from Ref.[85]. Copyright (2019) American Chemical Society.

## 3.5. Freestanding 2D monolayer vs materials on a substrate

Generally, most synthesized materials can be divided into (i) single exfoliated monolayer or (ii) 2D materials synthesized on a substrate. In recent years, there is also a growing class of non-vdW 2D materials (i.e., compounds that are usually stabilized in the form of a couple of layers of 2D materials due to their ability to reconstruct dangling bonds or strong chemical passivation due to a reactive surface). While non-vdW 2D materials are not discussed in detail within this paper, the main conclusions remain the same. What makes the different types of materials unique is the fact that if the 2D material is isolated in the freestanding form, its properties can indeed be well analyzed within DFT and even potential device performance can be predicted. However, for the class of materials synthesized on a substrate (i.e., that cannot be isolated as a freestanding single layer), theoretically predicted phenomena in freestanding monolayer systems may not be realized experimentally. For instance, let us consider silicene. According to DFT calculations, the freestanding monolayer of silicene has a zero band gap with the Dirac cone at the Fermi level.[91] However, experimental synthesis of silicene on various substrates like Ag(111)[106], Ag(110)[107] , ZrB2(0001)[108] , and Ir(111)[109] does not agree with the theoretical predictions.[110] The substrate/material interface can drastically modify the electronic, optical, and magnetic properties of deposited compounds. Importantly, the substrate effect does not only depend on the specific properties of the compound such as polarity, thickness, stoichiometry, and surface dangling bonds, but also depends on the substrate properties which can be inorganic or organic, polar or non-polar, and normal-insulator, metal or topological insulator. In general, the substrate effects that modify the theoretically predicted properties in 2D compounds are classified as (i) strain[111], (ii) charge transfer[112], (iii) dielectric screening[113,114] , and (iv) optical interference.[115]

***(i) Strain***: Owing to interface formation, 2D material can exhibit the strain effect, which can be sufficiently large to change the original properties of 2D material as shown in Fig. 9. Moreover, not all strains are actually allowed in a 2D compound on top of a substrate, and hence, the theoretical prediction of exotic properties (e.g., topological symmetry protection and superconductivity) in high strained materials is not easily achieved.[116]

***(ii) Charge transfer:*** When a monolayer 2D semiconductor is located on a substrate, it can exhibit direct charge transfer or charge density redistribution[117], which is sufficient to affect 2D semiconductors. For example, using DFT calculations for monolayer silicene on different Ir, Cu, Mg, Au, Pt, Al, and Ag substrates Quhe et. al., showed the charge transferred between the substrate and the 2D material which can destroy the exotic properties of silicene.[8]

***(iii) Dielectric screening***: The dielectric mismatch between a nanoscale semiconducting material and the surrounding environment can result in a number of peculiarities that are not present in bulk.[118] For example, the dielectric environment can change the exciton binding energy in semiconducting films or nanowires by screening Coulomb interactions and modifying electron-hole interactions[119] . Thus it has a crucial influence on optical bandgaps of semiconductors and can strongly modulate both electrical and optical properties of low dimensional materials, including quantum wells[120], carbon nanotubes[121], and graphene.[122]

***Optical interference***: Optical interference, resulting from reflection and refraction at each interface of thin films and substrate, which can enhance or reduce the local electric fields (E-fields) around the surface depending on optical constants and film thicknesses. It means that the amplitude of local E-fields can be modulated by the optical interference effect. This effect must be considered when a material is ultrathin as in this case the properties of the material are very sensitive to the local E-fields around the surface.[123] For example, intensities of Raman signals for silicene strongly depend on both incident local E-field and emitted local E-field.[124]

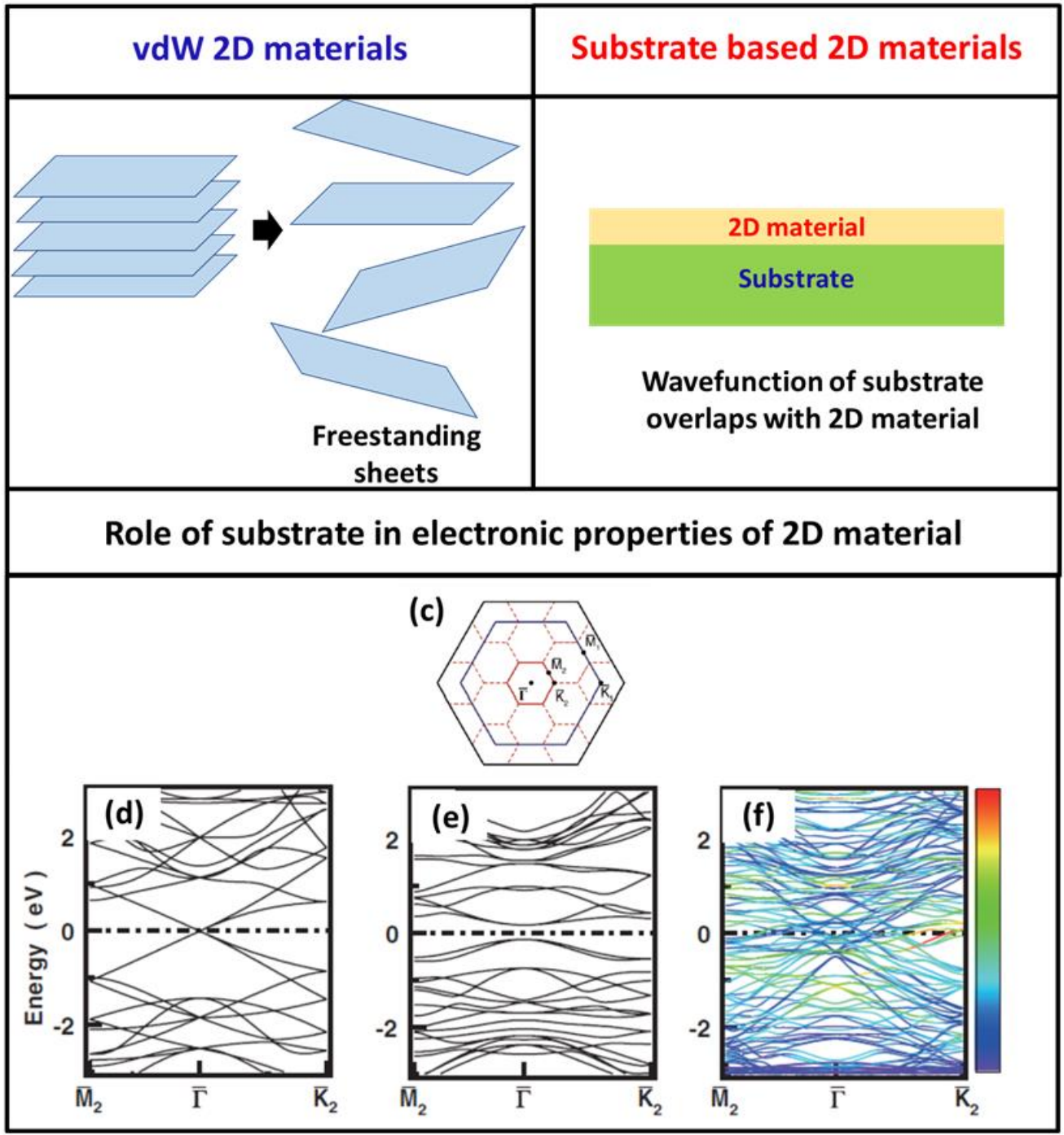

**Figure 9**. Schematic diagram of (a) vdW and (b) substrate based 2D materials. (c) Brillouin zone corresponding to silicene unit cell. Electronic band structure of (d) free standing silicene (e) distorted silicene and (f) 4x4 silicene on Ag(111) substrate. In Fig (f), color bar shows the higher (red) to lower (blue) contribution from Si $p_z$ orbital. Lower panel of the figures (c-f) are reproduced from ref[125] Copyright (2013) by The American Physical Society.

## 4: Future of first-principles studies of 2D materials

With the development of electronic structure theory, it becomes clear that the properties of many compounds can be well described within DFT as long as the proper atomic identities, composition, and structure are used. What makes the case of theoretical studies of 2D materials special is the fact that the field is relatively immature – there are no universal standards for the prediction of the material properties and their realizability. This led to the situation when the theoretical works often become similar to a Lego game where scientists often do not use the important criteria for validation of their results or even use methods that should not be used to describe corresponding materials properties. While it is only a matter of time until clear guidance is developed for the field, herein, we foresee two main research challenges:

**(i)**      **Development of advanced methods to analyze properties of 2D materials:** Description of 2D materials as periodic solids with a large thickness of the vacuum does not allow to directly apply every theory developed for the bulk systems towards the 2D materials. This is demonstrated above on the example of optical properties of 2D materials, but the problem is not limited to that. For instance, charged defects cannot be described within the standard defect corrections developed for bulk systems.[126,127] This means that DFT should not be used as a black box method for 2D compounds as this can eventually increase the gap between theory and experiment, leading to significant contradiction and misunderstanding in the field.

**(ii)**      **Validation of the reliability of the theoretical predictions:** The 2D community is largely driven by applied science resulting in the prediction of materials properties for specific applications or even the expected device performance. The demonstration of the fundamental science does not always require using the compounds that can be realized, while in the case of applied science, one should have clearly developed criteria to be satisfied to propose the specific application. Such experience should, for instance, include the criteria for the realization of 2D materials and validation of their performance. This may be limited to the realization criteria discussed above or, more generally, by developing a theory of stable/metastable materials predicting the conditions under which materials can be realized.

We believe that by keeping in mind the main limitations indicated in the current work, the theoretical 2D community can substantially reduce the gap between the existing theoretical research and the experimental works.

**Acknowledgment:**

The authors thank ENSEMBLE3 Project (MAB/2020/14) which is carried out within the International Research Agendas Programme (IRAP) of the Foundation for Polish Science co-financed by the European Union under the European Regional Development Fund and Teaming Horizon 2020 (GA no. 857543) programme of the European Commission. GDM thanks financial support from Brazilian funding agencies FAPESP and CNPq. This research was supported in part by PLGrid infrastructure.